\documentclass[]{raa}            

\usepackage{graphicx,times}             
\usepackage{natbib}
\usepackage{amssymb,amsmath}
\bibpunct{(}{)}{;}{a}{}{,}

\usepackage[a4paper=true,pagebackref=true]{hyperref}
\hypersetup{colorlinks = true, linkcolor = green, anchorcolor = red, citecolor = blue, filecolor = red, pagecolor = red, urlcolor = red}
\begin{document}

   \title{The Gravitational Wave Emission of Double White Dwarf Coalescences
}

   \volnopage{Vol.0 (20xx) No.0, 000--000}      
   \setcounter{page}{1}          

   \author{Ze-Cheng Zou
      \inst{1}
   \and Xiao-Long Zhou
      \inst{1}
   \and Yong-Feng Huang
      \inst{1,2}
   }

   \institute{School of Astronomy and Space Science, Nanjing University,
   Nanjing 210023, P. R. China; \emph{hyf@nju.edu.cn}\\
        \and
        Key Laboratory of Modern Astronomy and Astrophysics (Nanjing University),
        Ministry of Education, P. R. China\\
\vs\no
   {\small Received~~20xx month day; accepted~~20xx~~month day}}

\abstract{ Type Ia Supernovae (SNe Ia) are widely used as standard candles to probe the Universe. However, 
          how these fierce explosions are produced itself is still a highly debated issue.  
          There are mainly two popular models for SNe Ia, i.e. the
          double-degenerate scenario and the single-degenerate scenario. The double-degenerate scenario
          suggests that SNe Ia are produced by the coalescence of two degenerate white dwarfs, 
          while the single-degenerate scenario suggests that the continuous accretion of a single degenerate 
          white dwarf from its normal stellar companion will finally lead to a disastrous explosion when  
          it is over-massive,     
          resulting in an SN Ia. The rapid development of the gravitational wave astronomy sheds new 
          light on the nature of SNe Ia. In this study, we calculate the gravitational wave emissions
          of double white dwarf coalescences and compare them with the sensitivities of several upcoming
          detectors. It is found that the gravitational wave emissions from double white dwarf mergers 
          in the locale universe are strong enough to be detected by LISA. We argue that LISA-like gravitational 
          wave detectors sensitive in the frequency range of 0.01 --- 0.1 Hz will be a powerful tool to 
          test the double-degenerate model of SNe Ia, and also to probe the Universe. 
\keywords{White dwarfs, Gravitational waves, Supernovae: general}
}

   \authorrunning{Z.-C. Zou, X.-L. Zhou \& Y.-F. Huang }            
   \titlerunning{The Gravitational Wave Emission of Double White Dwarf Coalescences}  

   \maketitle

%
%
\section{Introduction}           
\label{sect:intro}

Type Ia supernovae (SNe Ia) represent a kind of high-energy astrophysical phenomena with similar light curves
in the universe. Typically, the peak luminosity of SN Ia, the color at the peak time, and the decline rate of
the brightness after the peak luminosity are found to follow the so-called Phillips relationship
(\citealt{Phillips+1993,Phillips+etal+1999,Kattner+etal+2012}). This makes SNe Ia a useful tool to act
as standard candles in cosmological distance measurements. They have been widely used in determining cosmological
parameters. Using SNe Ia, researchers have found the acceleration of cosmic
expansion (\citealt{Perlmutter+etal+1999,Riess+etal+1998}). However, note that the explosion mechanism of SNe Ia and
their exact progenitors are still highly debated (\citealt{Nomoto+etal+1997,Wang+Han+2012,Maoz+etal+2014}).

In principle, SNe Ia are believed to be related to electron-degenerate matter. When an electron-degenerate star,
i.e. a white dwarf (WD), accretes continuously from its companion, it will finally become more massive than
the Chandrasekhar limit and the electron-degenerate pressure will not be able to balance the self-gravity.
It will result in an explosive collapse and give birth to an SN Ia. Still, there are at least two popular scenarios
of SNe Ia following this idea, i.e., the single-degenerate scenario and the double-degenerate
scenario. In the single-degenerate scenario, a WD in a binary system accretes matter from its normal companion star.
The accretion continues till the WD reaches the Chandrasekhar limit and the explosion happens 
(\citealt{Hachisu+etal+1996,Li+vandenHeuvel+1997,Han+Podsiadlowski+2004,Wang+etal+2009}). In the double-degenerate
scenario, two WDs form a binary system and lose their orbital energy due to gravitational wave (GW) emission.
At the last stage of the inspiraling, the less massive WD is tidally disrupted and a merger event is subsequently
produce. An SN Ia will occur if the merger leads to central carbon ignition (e.g., \citealt{Iben+Tutukov+1984,Webbink+1984}).

The single-degenerate scenario is more popular nowadays, yet it faces many serious problems (e.g., \citealt{Wang+2018}).
For example, the details of the detonation waves and the ignition are still quite unclear (\citealt{Peng+1999}).
Additionally, there is a lack of hydrogen in the observed spectra of SNe Ia (\citealt{Livio+Mazzali+2018}).
The double-degenerate scenario, on the contrary, is well consistent with this spectral feature. Moreover, the existence
of a significant amount of double WD systems have been directly proved by various
observations (\citealt{Iben+Tutukov+1984,Iben+Livio+1993,Saffer+etal+1998,Roelofs+etal+2010}), and some of them
have even been found to show orbital decay as predicted by the general relativity theory so that they should definitely merge
in the future (e.g., \citealt{Brown+etal+2011,Hermes+etal+2012,Kilic+etal+2014}). More interestingly,
several authors recently calculated the chemical outcome of double WD collisions by means of numerical simulations.
They found that their results are consistent with observed SN Ia spectra
(\citealt{Kushnir+etal+2013,Dong+etal+2015,Isern+Brav+2018}). 
Meanwhile, we should also note that the double-degenerate scenario also faces some serious problems. 
First, it has difficulties in explaining the similarities of most SNe Ia as the WD explosion mass has a 
relatively wide range. Second, several studies reveal that the outcome of double WD mergers 
may be a neutron star resulting from an accretion-induced collapse, rather than a 
thermonuclear explosion as required by the observed SNe Ia (\citealt{Nomoto+Iben+1985,Saio+Nomoto+1998}). 
These arguments are based on the assumption that the merging remnant consists of a hot envelope or a thick 
disc, or even both upon the primary WD. In this case, the accreting oxygen-neon WD  
would finally collapse into a neutron star when it approaches the Chandrasekhar limit. 
Interestingly, \cite{Pakmor+etal+2010} proposed a new violent merger scenario for SNe Ia based on the merging 
of double WDs. In their scenario, a prompt detonation is triggered while the merger is still ongoing, 
giving rise to an SN Ia explosion. In short, the final fates (collapse to a neutron star or thermonuclear 
explosion as an SN Ia) of double WD mergers are strongly dependent on the merging processes 
(e.g. slow merger, fast merger, composite merger, violent merger, etc). 

While the trigger mechanism of SNe Ia is still quite unclear, the rapid development of GW
astronomy could shed new light on their nature. In the double-degenerate scenario, a strong GW emission is expected
as the two degenerate WD stars spiral in before the final merging. Decihertz interferometers such as the DECIGO
and B-DECIGO (\citealt{Isoyama+etal+2018,Sato+etal+2017,Yagi+Seto+2011}) can detect the most massive binary
WDs in our Galaxy (\citealt{Maselli+etal+2019}), and less massive binary WDs may be targets for detectors
working in millihertz regime like LISA. On the contrary, in the single-degenerate scenario,
GW emission should be weak during the whole process.

In this study, we investigate the last stage of double WDs' inspiraling and their GW emissions. We simulate the
evolution of the GW signals and obtain the spectra features. Especially, we examine whether LISA can detect
the GW signature if SNe Ia are produced by double-degenerate star mergers.
The structure of our paper is organized as follows. In Sect. \ref{sect:StraAmp}, we describe the strain amplitude
of GW from double WD systems. In Sect. \ref{sect:StraSpec}, the observable quantity of strain spectral amplitude is
introduced. In Sect. \ref{sect:Cutoff}, the cut-off frequency of the gravitational wave from a merging double WD
system is calculated and compared with the spectral range of GW detectors.
In Sect. \ref{sect:num}, we calculate the GW emission produced by some recent SNe Ia in framework of the
double-degenerate scenario, and confront the results with GW detectors.
Finally, our conclusions and discussion are presented in Sect. \ref{sect:conclusion}.


\section{Gravitational Wave Strain Amplitude}
\label{sect:StraAmp}

In this study, we consider a binary system consisting of two WDs with equal mass $M$ rotating in a
circular orbit around the common barycentre. It has been argued that in the double-degenerate scenario,
the mass ratio (defined as the mass of the less massive WD divided by the mass of its more massive
companion) of the two WDs should be larger than 0.8 to ignite an SN Ia (\citealt{Pakmor+etal+2011}).
So, the equal-mass assumption is acceptable in considering the GW emission from the system.
Since the separation between the two WDs is still relatively large even at the final stage that
they are disrupted by tidal force, Newtonian mechanics is roughly applicable for the orbital motion
during the whole inspiraling process. According to Kepler's law, the orbit frequency is
\begin{equation}
  f_{orb}=\sqrt{\frac{GM}{2\pi^2R^3}},\label{eq:1}
\end{equation}
where $R$ is the separation between the two stars and $G$ is the gravitational constant.
Correspondingly, the frequency of GW is
\begin{equation}
  f=2f_{orb}=\sqrt{\frac{2GM}{\pi^2R^3}}.\label{eq:2}
\end{equation}
The GW power emitted by the binary can be calculated by the following formula
\begin{equation}
  P=\frac{64G^4M^5}{5c^5R^5},\label{eq:3}
\end{equation}
where $c$ is the velocity of light.

As the double WD system emitting GW at the aforementioned power, the frequency of
GW will gradually decline as (\citealt{Creighton+Anderson+2011})
\begin{equation}
  \dot{f}={\frac{96}{5}\frac{c^3}{G}\frac{f}{M_c}\left(\frac{G}{c^3}\pi f\ M_c\right)}^{8/3},\label{eq:4}
\end{equation}
due to the energy loss, where $M_c$ is the chirp mass defined as
\begin{equation}
  M_c=\frac{\left(m_1m_2\right)^{3/5}}{\left(m_1+m_2\right)^{1/5}}.\label{eq:5}
\end{equation}
The GW can have two polarization states, with the amplitude usually designated as $h_+$ and $h_\times$,
respectively (\citealt{Landau+Lifshitz+2012,Postnov+Yungelson+2014}). The overall GW amplitude is then
\begin{equation}
  h=\sqrt{\left(h_+\right)^2+\left(h_\times\right)^2}.\label{eq:6}
\end{equation}
After averaging the gravitational wave signal in whole period, the amplitude is
\begin{equation}
  h=\left(\frac{32}{5}\right)^2\frac{G}{c^2}\frac{M_c}{d}\left(\frac{G}{c^3}\pi f\ M_c\right)^{2/3},\label{eq:7}
\end{equation}
where $d$ is the distance of the GW source with respect to us (\citealt{Postnov+Yungelson+2014}).

\section{Gravitational Wave Strain Spectral Amplitude}
\label{sect:StraSpec}

In studying the detectability of GWs, it is more important to consider the signal-to-noise ratio (SNR),
i.e., whether the effect of GWs is higher than the sensitivity of the detector
(\citealt{Robson+etal+2019,Wong+etal+2018}). In the inspiraling phase, the energy density of GW
(\citealt{Postnov+Yungelson+2014}) can be expressed as
\begin{equation}
  S_h=\frac{G^{5/3}}{c^3}\frac{\pi}{12}\frac{M_c^{5/3}}{d^2}\frac{1}{\left(\pi f\right)^{7/3}}.\label{eq:8}
\end{equation}
The strain spectral amplitude ($h_f$) is the Fourier transformation of $h$.
In the inspiraling phase of a binary WD system, the frequency generally evolves very slowly,
so that the strain spectral amplitude can be expressed as (\citealt{Postnov+Yungelson+2014,Geng+etal+2015,Robson+etal+2019}),
\begin{equation}
  h_f=\sqrt{S_h}=\sqrt{\frac{G^{5/3}}{c^3}\frac{\pi}{12}\frac{M_c^{5/3}}{d^2}\frac{1}{\left(\pi f\right)^{7/3}}}.\label{eq:9}
\end{equation}
This equation is applicable mainly at the late stage of the inspiraling process. It is accurate enough
even at the moment when the less massive WD is to be disrupted by its companion.

However, note that at the early stage of the inspiraling, the two WDs are far from each other
and the variation rate of the GW frequency is extremely small, i.e. $\dot{f}\times\ T_{obs}<1/T_{obs}$,
where $T_{obs}$ is integration time for a continuous observation. It means that the emitted GW is nearly
monochromatic and the observed GW frequency almost does not change during the observation period of
$T_{obs}$. As a result, it is the integration time that restricts the observed SNR and
the above expression of $h_f$ is inappropriate. 
For a particular integration time $T_{obs}$, to determine when 
the expression of $h_f$  (i.e. Eq. \ref{eq:9}) begins to be applicable, we can set $\dot{f}\times\ T_{obs} = 1/T_{obs}$
and derive the corresponding GW frequency as
\begin{equation}
  f_T=\left(\frac{8}{3}\frac{\kappa}{T_{obs}^2}\right)^{3/11},\label{eq:10}
\end{equation}
where $\kappa$ is defined as
\begin{equation}
  \kappa=\frac{5}{256}\left(\frac{G}{c^3}M_c\right)^{-5/3}\pi^{-8/3}.\label{eq:11}
\end{equation}
When $f>f_T$, the evolution of the GW frequency becomes significant and the sensitivity are determined by 
Eq. \ref{eq:9}.  
In our study, we take the integration time as 4 years for all the following numerical calculations. 
This is the planed duty time of LISA L3 mission (\citealt{Amaro-Seoane+etal+2017}).

\section{Cut-off frequency of the Gravitational Wave}
\label{sect:Cutoff}

The inspiraling is a gradual process. With the gradual decrease of the orbital separation, the GW frequency
increases correspondingly. As a result, according to Eqs. \ref{eq:7} and \ref{eq:9}, the strain amplitude
of GW increases while the strain spectral amplitude decreases gradually. Finally, the GW emission will be
ceased when the less massive WD is tidally disrupted by its companion. This happens at the so called
tidal disruption radius (\citealt{Hills+1975}), when the separation between the two WDs is
\begin{equation}
  R_{td}\approx\left(\frac{6M}{\pi\rho}\right)^{1/3},\label{eq:12}
\end{equation}
where $\rho$ is the mean density of the less massive white dwarf.
Taken $\rho = {10}^9\,{\rm kg\,m^{-3}}$, we get $R_{td} \approx 1.56\times 10^7{(M/M_\odot)}^{1/3}\rm m$.
At this separation, the gravity is $1.08\times 10^{37}{(M/M_\odot)}^{4/3}\rm N$. It means that the gravity is not
too strong and the Newtonian mechanical assumption is applicable.

Just before the tidal disruption separation, the GW emission power comes to a maximum and the GW frequency
is also the highest. After being tidally disrupted, the GW emission will be almost completely cut-off.
Combining Eqs. \ref{eq:2} and \ref{eq:12}, the cut-off frequency can be derived as
\begin{equation}
  f_{cut}=\sqrt{\frac{G}{3\pi}\rho}.\label{eq:13}
\end{equation}
A rough estimation shows that the cut-off frequency of GW from binary WDs are around 1\,Hz.
Especially, at this last stage, the variation rate of the GW frequency is around $10^{-5}\,\mathrm{s}^{-2}$.
We see that this rate is actually small enough so that the applicability of Eq.~\ref{eq:9} is guaranteed.

\section{Numerical Results}
\label{sect:num}

\begin{figure}[t]
  \captionsetup{margin=1em}
  \begin{minipage}[t]{.5\textwidth}
    \centering
    \includegraphics[width=\textwidth, angle=0]{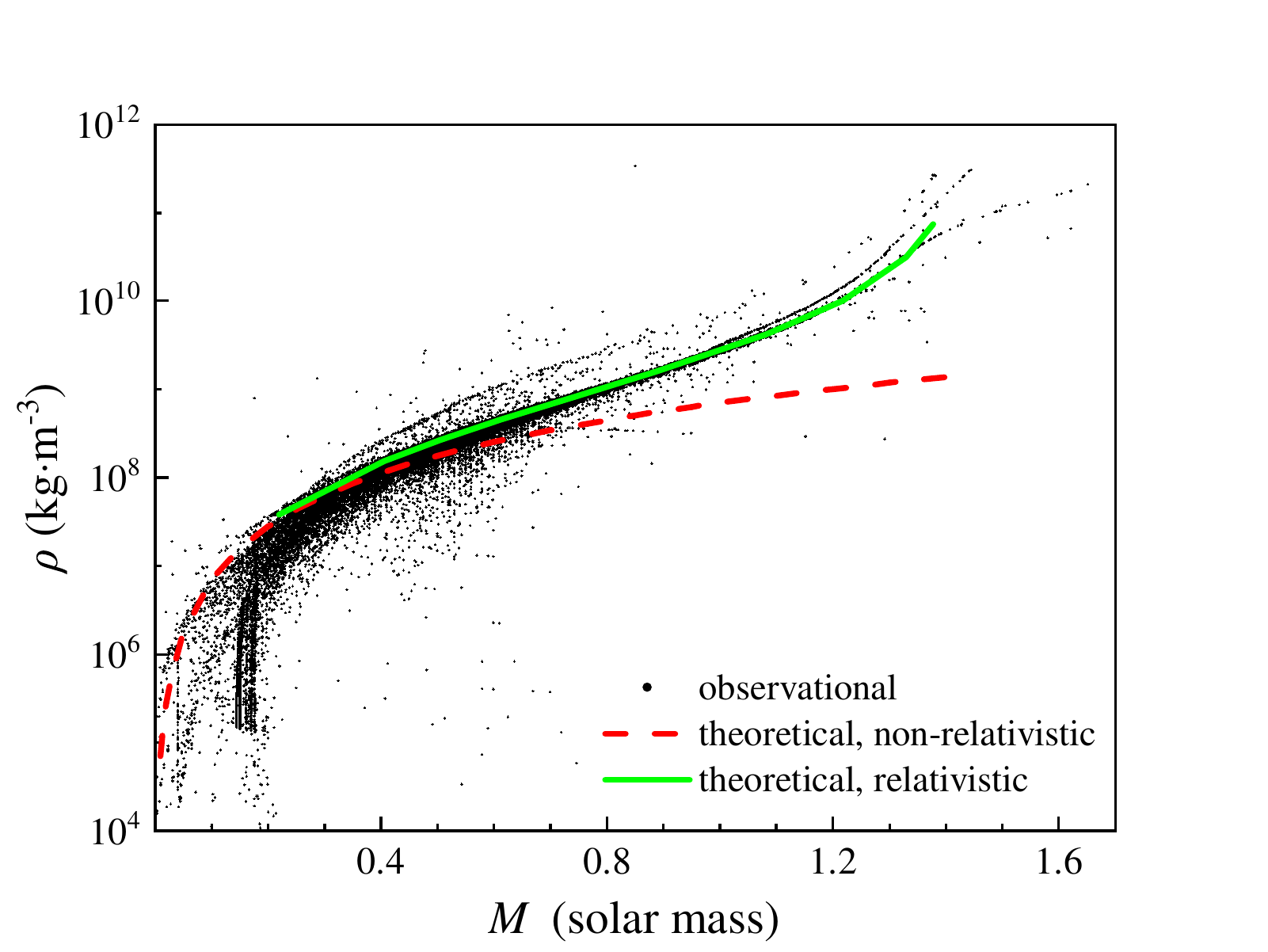}
    \caption{Mean density versus mass for the currently observed WDs. Each point represents a WD.
    The observational data are
    taken from the Montreal White Dwarf Database (\citealt{Dufour+etal+2017}). The mean density
    is calculated from the observed mass and the surface gravitational acceleration,
    assuming a spherical configuration without spinning. The dashed curve and the solid 
    curve represent theoretical non-relativistic (\citealt{Chandrasekhar+1935}) and 
    relativistic (\citealt{Chandrasekhar+1967}) mass-density relations of WDs, respectively.}
    \label{Fig1}
  \end{minipage}%
  \begin{minipage}[t]{.5\textwidth}
    \centering
    \includegraphics[width=\textwidth, angle=0]{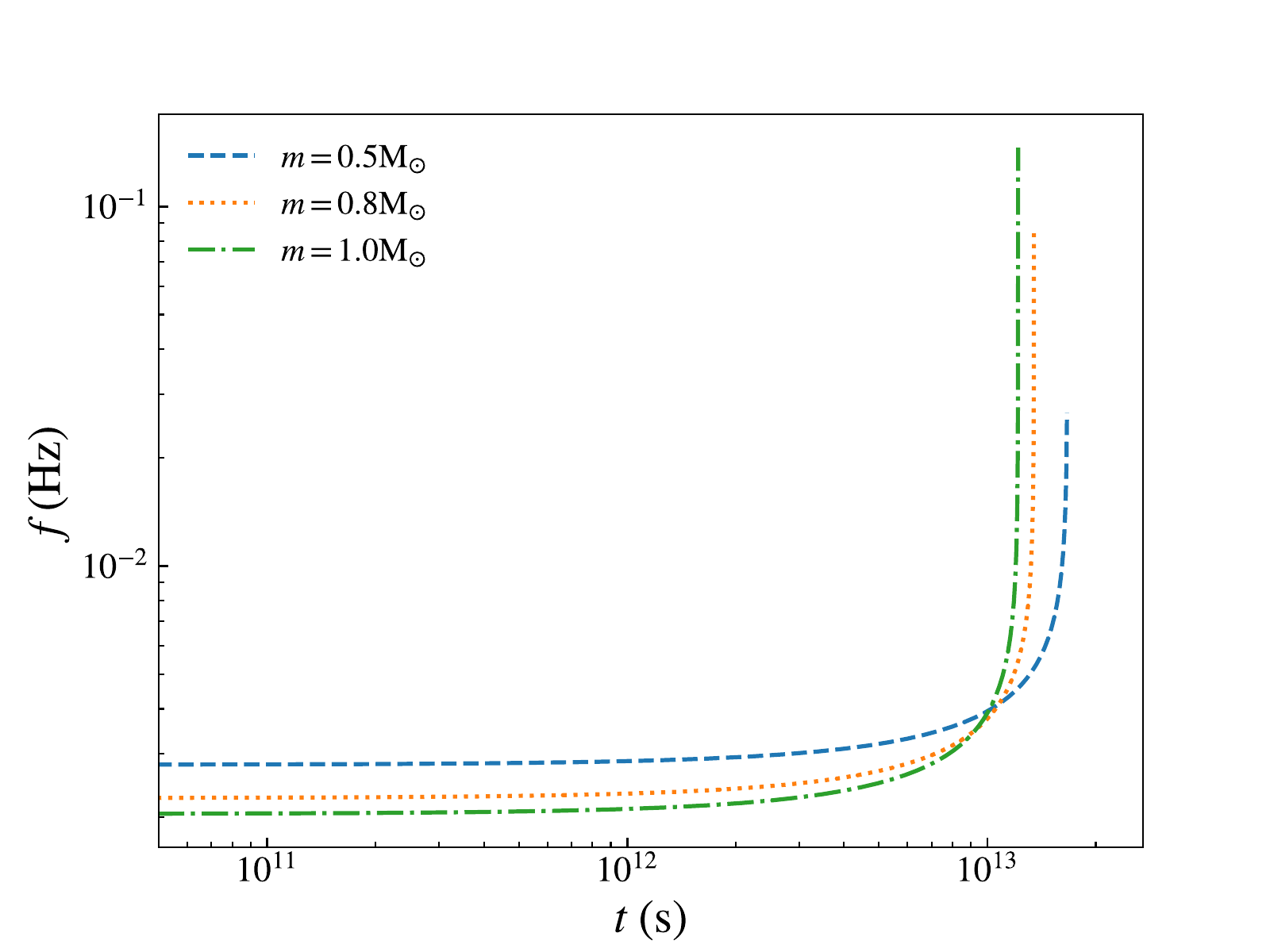}
    \caption{Evolution of the GW frequency for a double WD system. The orbit is assumed to be circular
     throughout the inspiraling process. $t$ = 0 corresponds to the moment that the GW frequency equals to
     $f_T$. The dashed, dotted, and dash-dotted lines correspond to a WD mass of $0.5 M_\odot$, $0.8 M_\odot$
     and $1.0 M_\odot$, respectively. For each curve, the highest frequency corresponds to
     the cut-off frequency when the WDs are tidally disrupted. }
    \label{Fig2}
  \end{minipage}
\end{figure}

According to Eqs. \ref{eq:7} and \ref{eq:9}, the strain amplitude and the strain spectral amplitude
of GWs are mainly determined by the masses of the WDs. In this study, for simplicity, we assume that the two
companion WDs are equal in mass. 
As an exemplar investigation, we take three typical values for 
the WD mass in our calculations, i.e. $0.5\,M_\odot$, $0.8\,M_\odot$, and $1.0\,M_\odot$.  
Note that the $0.5\,M_\odot$-WD binary mergers in fact are unlikely lead to SNe Ia.  
In the violent merger scenario, a detonation leading to an SN Ia explosion in the merger phase may not be 
triggered for double WDs with individual mass lower than $0.8\,M_\odot$ (\citealt{Sim+etal+2010}). Moreover, 
\cite{Sato+etal+2015} also argued that for the primary WD of $0.7\,M_\odot<M<0.9\,M_\odot$, a total mass 
larger than $1.38\,M_\odot$ is required to trigger an SN Ia detonation in the stationary rotating 
merger remnant phase. However, since low-mass WDs are very common (see Fig.~\ref{Fig1}) and low-mass binary 
WDs form the most numerous population among all binary WDs (\citealt{Postnov+Yungelson+2014}), we use the 
$0.5\,M_\odot$ case as a representation of these interesting low mass WD binaries. In fact, they are 
important goals of many GW experiments.

At the same time, we notice that the cut-off frequency of the GW emission mainly depends on the density 
of WDs (see Eq.~\ref{eq:13}). In order to calculate the GW evolution of binary WDs of certain masses, 
we need to know the relationship between density and mass of WDs. 
For this purpose, we mainly resort to the observational data of WDs.
The Montreal White Dwarf Database (MWDD) provides a useful data source, which lists the observed 
parameters for WDs (\citealt{Dufour+etal+2017}, http://montrealwhitedwarfdatabase.org/),
including the mass and radius for a significant portion of them so that the mean density can be derived.
Using the MWDD data, we have plotted in Fig.~\ref{Fig1} the density versus mass for all the observed WDs with data available.

Inside a WD, the gravity is balanced mainly by the pressure of degenerate electrons. 
So the mass-radius relation of WDs can be conveniently derived by theoretical analysis. A simple 
dimensional analysis gives that the radius of a WD with mass $M$ scales as 
$ r \propto \mu_e^{-5/3} M^{-1/3} $, where $\mu_e$ is the average baryon number per 
electron (\citealt{Koester+Chanmugam+1990}). In fact, in the non-relativistic case, 
the detailed mass-radius relation has been derived by \cite{Chandrasekhar+1935} as, 
\begin{equation}
  r \approx 2.785 \times 10^9 {\rm cm} \; \times (M/1 M_\odot)^{-1/3} {\mu_e}^{-5/3}. \label{eq:17}
\end{equation}
Note that Eq.~\ref{eq:17} is applicable mainly for low mass WDs, which have relatively larger 
radii so that the gravity on the WD surface is not too strong. For higher mass WDs, 
the relativistic effect should be taken into consideration. In this case, the equilibrium 
equations become much more complicated and could only be numerically solved. Numerical 
results on relativistic WDs have also been obtained and provided as a data table 
by \cite{Chandrasekhar+1967}. To present a direct comparison with observations, we have 
plotted the theoretical mass-density relations of WDs in Fig.~\ref{Fig1}. The dashed curve is 
the non-relativistic relation calculated from Eq. \ref{eq:17}, taking $\mu_e=2$, and the solid curve 
corresponds to the relativistic relation as presented in the data table of \cite{Chandrasekhar+1967}. 
We see that the non-relativistic curve matches well with the observational data points 
in the low mass segment, while the relativistic curve is consistent with the data points 
even at the high mass end. However, we should also note that the observational data points 
are generally quite scattered as compared with the theoretical curve, 
which means WDs of a particular mass can have very different radii. This may be caused by their 
different internal composition. Also, the crust may play an important role on the WD radius. 

\begin{figure}[!h]
  \captionsetup{margin=1em}
  \begin{minipage}[t]{.5\textwidth}
    \centering
    \includegraphics[width=\textwidth, angle=0]{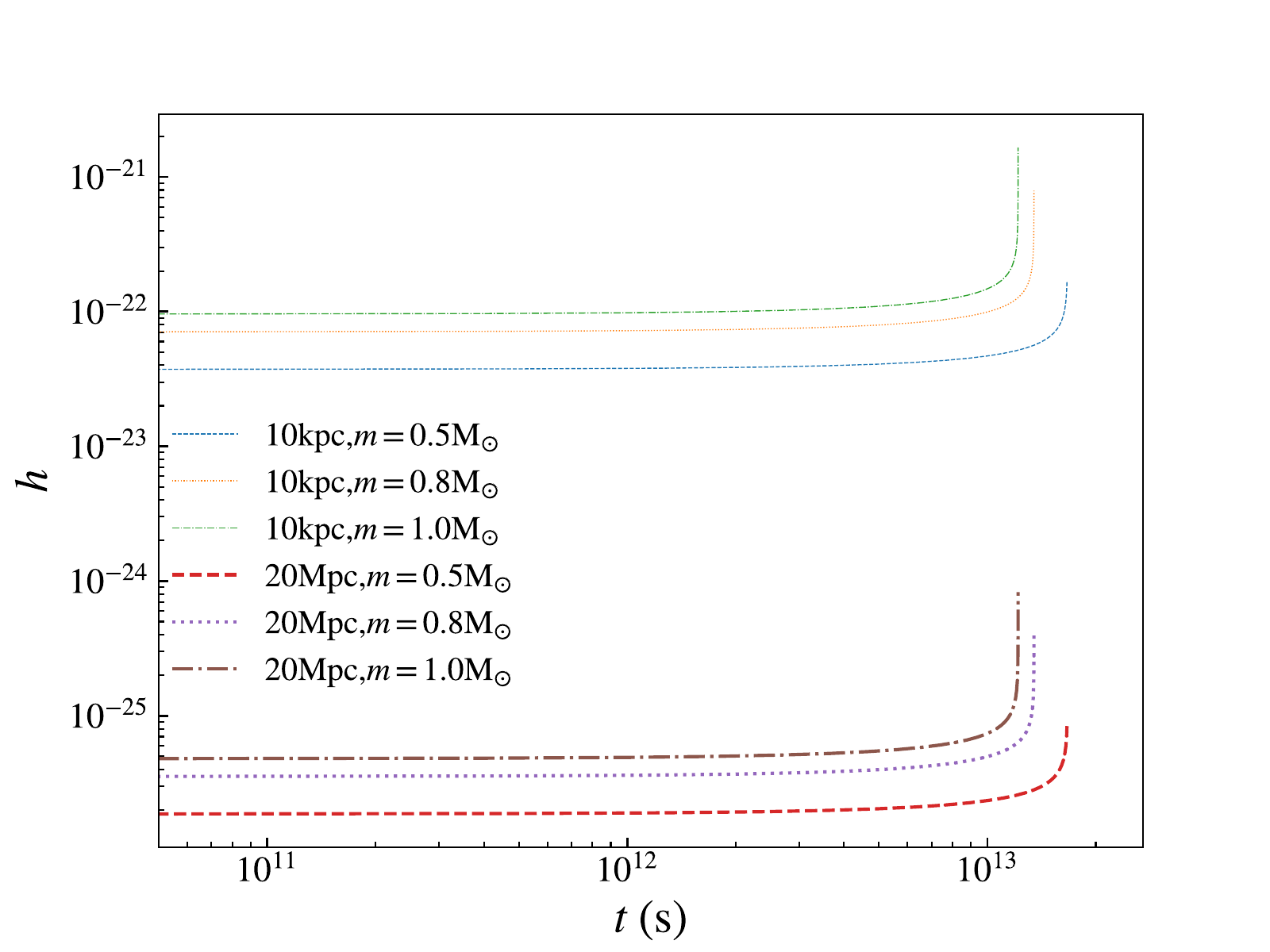}
    \caption{Evolution of the GW strain amplitude for double WD systems. The parameters
     are the same as in Fig. \ref{Fig2}. The thin dashed, dotted, and dash-dotted lines corresponds to
     a WD mass of $0.5 M_\odot$, $0.8 M_\odot$, and $1.0 M_\odot$, respectively, at a distance of
     10 kpc. For the thick lines, the distance is taken as 20 Mpc while other parameters remain unchanged. }
    \label{Fig3}
  \end{minipage}%
  \begin{minipage}[t]{.5\textwidth}
    \centering
    \includegraphics[width=\textwidth, angle=0]{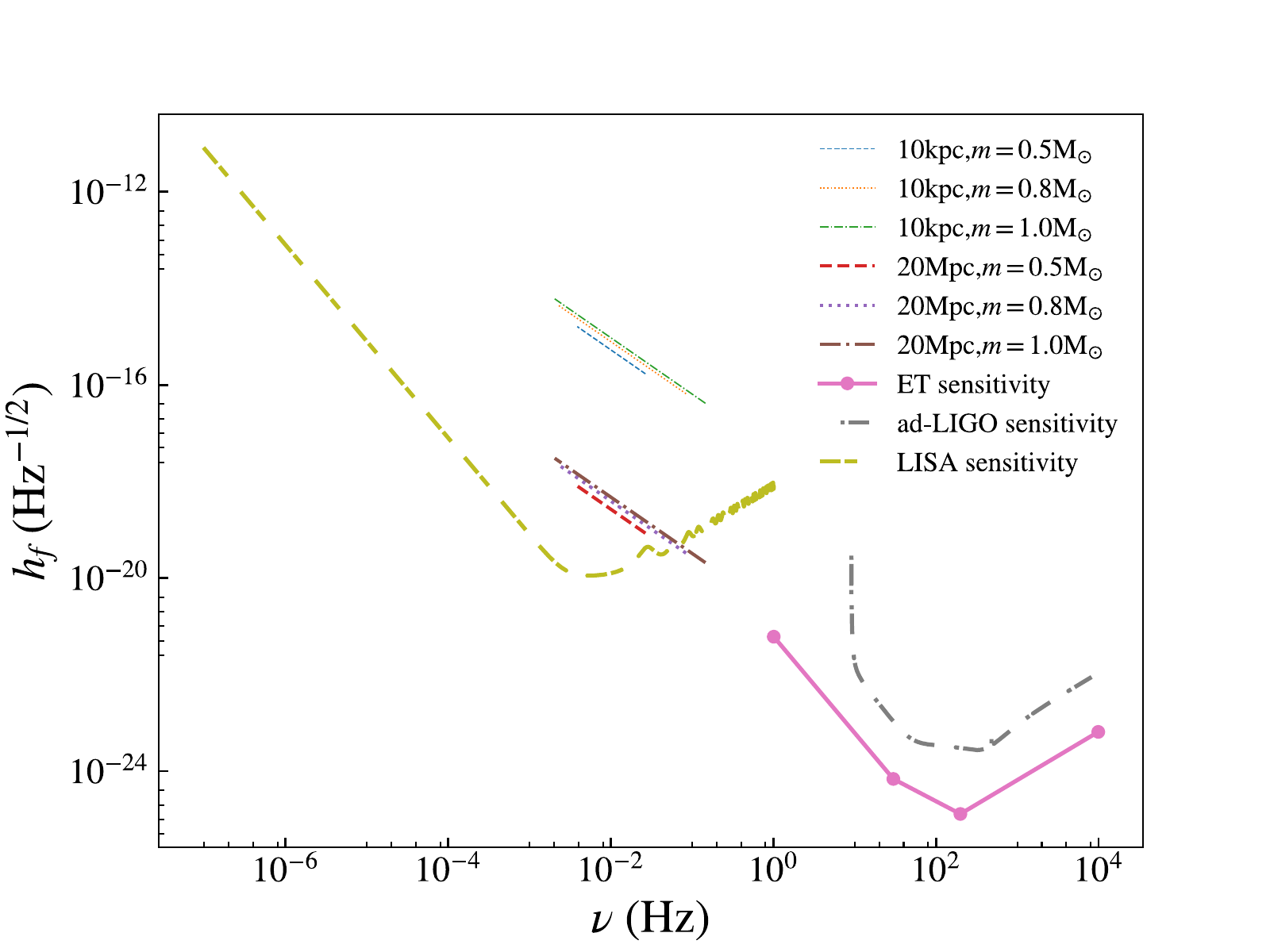}
    \caption{ Evolution of the GW strain spectral amplitude for double WD systems.
    The parameters and line styles are the same as in Fig.~\ref{Fig3}.
    As a comparison, the sensitivity curves of Einstein Telescope (the thick solid curve;
     \citealt{Hild+etal+2008}), advanced LIGO (the thick dash-dotted curve; \citealt{Harry+2010}), and LISA
     (the thick double-dashed curve; \citealt{Larson+etal+2000}) are also plotted. }
    \label{Fig4}
  \end{minipage}
\end{figure}

From Fig.~\ref{Fig1}, we can clearly see that more massive WDs tend to have a higher 
mean density. The observed data points could provide a reliable clue on the WD density.
Especially, we see that when the WD mass is $0.5\,M_\odot$, $0.8\,M_\odot$, or $1.0\,M_\odot$,
the corresponding mean density is typically ${10}^8\,{\rm kg\,m^{-3}}$, ${10}^9\,{\rm kg\,m^{-3}}$,
and $3\times{10}^9\,{\rm kg\,m^{-3}}$. With the hints from Fig.~\ref{Fig1}, we finally take the following 
three typical mass-density pairs in our subsequent calculations: 
$m_1=0.5\,M_\odot,\rho_1={10}^8\,{\rm kg\,m^{-3}};
m_2=0.8\,M_\odot,\rho_2={10}^9\,{\rm kg\,m^{-3}};
m_3=1.0\,M_\odot,\rho_3=3\times{10}^9\,{\rm kg\,m^{-3}}$.

Using the aforementioned typical parameters, we have simulated the dynamical evolution of the
inspiraling process of various double WD systems. In our calculations, we follow the inspiraling
process till the two WDs come to the tidal disruption radius so that the GW emission essentially
ceases. Fig. \ref{Fig2} illustrates the GW frequency versus time for the three cases. It can be
clearly seen that at the final stage, the GW frequency is typically in a range of 0.01 --- 0.1 Hz.
It is undetectable for LIGO and Einstein Telescope (which operate for GWs ranging from hertz to kilohertz),
but is well within the sensitive range of the future LISA experiment (from millihertz to hertz).
These GW events may also be targets of the B-DECIGO detector, as suggested by \cite{Maselli+etal+2019}.
Also, we see that when the WD mass is larger, the cut-off frequency is also significantly higher.
This is because for a high-mass WD, the mean density is also relatively high. Then according
to Eq.~\ref{eq:12}, the WD will be disrupted at a smaller radius, and the cut-off frequency
is correspondingly higher as indicated by Eq.~\ref{eq:13}.

In Fig.~\ref{Fig3}, we illustrate the evolution of the GW strain amplitude for double
WD systems. In this plot, we assume two different luminosity distances for the GW sources, 
10 kpc and 20 Mpc. Again we see that the GW strain amplitude is markedly enhanced when 
the WD mass increases. When the WD mass varies from 0.5 $M_\odot$ to 1.0 $M_\odot$ and 
increases by a factor of only two, the strain amplitude increases by a factor of about 10. 
In the 10 kpc distance cases, the strain amplitude can be as high 
as $10^{-22}$ --- $10^{-21}$ at the final stage. The amplitude is in the range of 
$10^{-25}$ --- $10^{-24}$ even in the 20 Mpc cases. 

\begin{figure}
  \centering
  \includegraphics[width=\textwidth, angle=0]{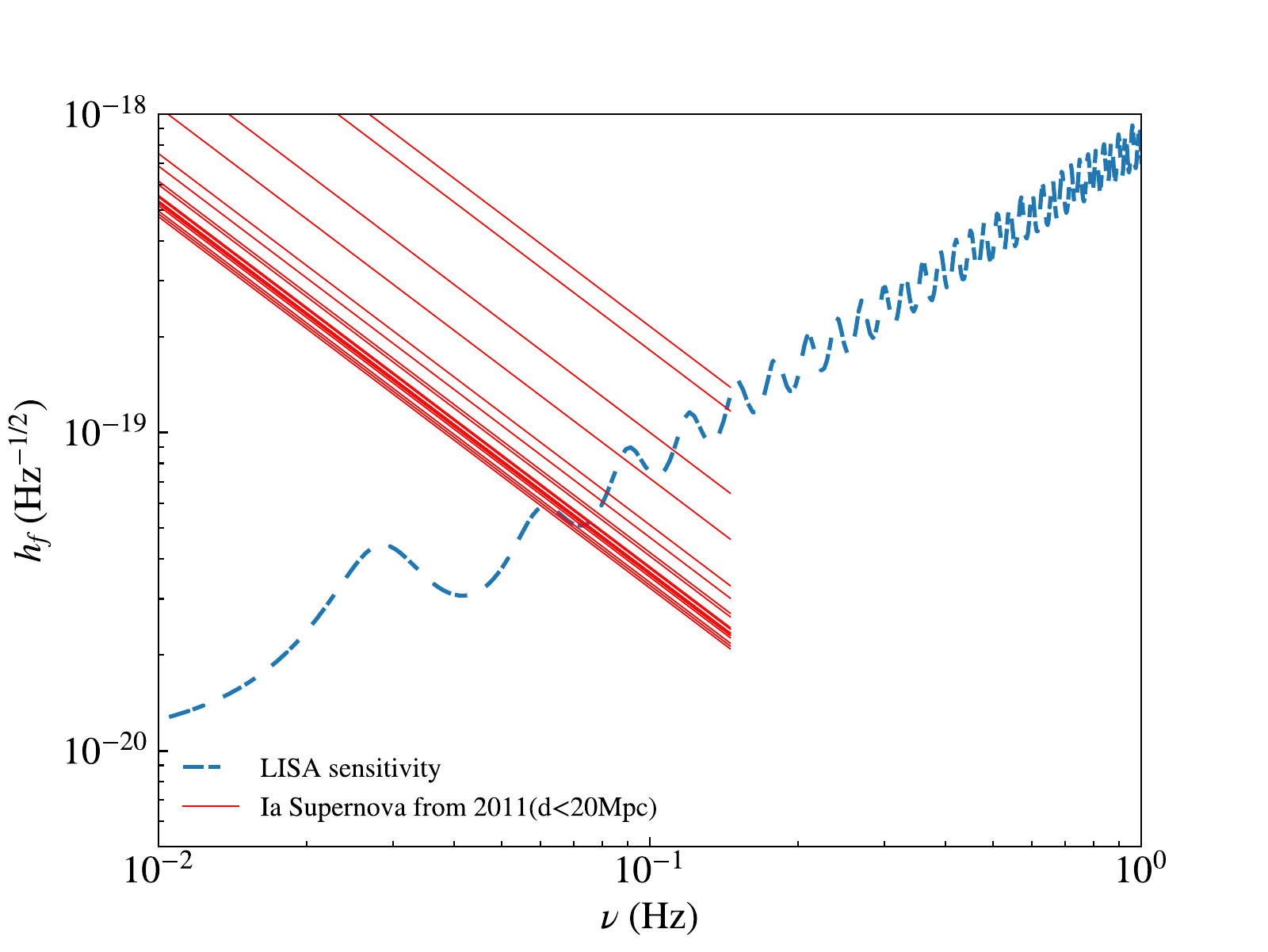}
  \caption{ Simulated GW strain spectral amplitude for some observed SNe Ia in the framework of
  the double-degenerate scenario. Each SN Ia is represented by a solid straight line.
  The dashed curve represents the sensitivity curve of LISA. }
  \label{Fig5}
\end{figure}

In order to know whether the GW is strong enough to be detected by GW instruments, 
we need to compare the strain spectral amplitude with the sensitivities of the detectors. 
In Fig.~\ref{Fig4}, we have plotted the strain spectra amplitude of different WD binary
systems, together with the sensitivity curves of LISA, advanced LIGO and Einstein Telescope. 
We see that GW emissions from double WD systems are not in the sensitive frequency ranges 
of advanced LIGO and Einstein Telescope, but are satisfactory targets for LISA.  
LISA can detect the GW of the binary WDs in our Milky Way Galaxy at a very high 
SNR (represented by the 10-kpc cases). More encouragingly, it can also detect such 
events in our local Universe to a distance much larger than 20 Mpc. 

Many SNe Ia have been observed till now. They are widely used in cosmology researches. If SNe Ia 
are produced via the double-degenerate mechanism, then the associated GW emissions should be 
detectable to LISA and the trigger mechanism can be firmed tested.  
The Open Supernova Catalog (\citealt{Guillochon+etal+2017}, https://sne.space/) provides a complete list of SNe Ia observed 
so far.  Using this data base, we have selected all the SNe Ia discovered
since 2011, with the luminosity distance smaller than 20\,Mpc. There are 19 events satisfying this criterion. 
We assume that all these SNe Ia were produced via the double-degenerate mechanism. For simplicity, the 
masses of the WDs are taken as 1.0 $M_\odot$ for all of them. In Fig.~\ref{Fig5}, the strain spectral 
amplitudes expected from these 19 events are plotted and compared with the sensitivity curve of LISA.   
Encouragingly, we see that all these events are detectable to LISA even in the earlier inspiraling stage. 
We thus argue that LISA will be a powerful tool to test the double-degenerate mechanism for SNe Ia.

\section{Conclusions and discussions}
\label{sect:conclusion}

Whether SNe Ia are produced by the single-degenerate mechanism or the double-degenerate mechanism 
is still a highly debated issue. In this study, we investigate the GW emissions from inspiraling double WD systems. 
Different from merging double black holes or double neutron stars (e.g. \citealt{Abbott+etal+2016,Abbott+etal+2017}), 
merging double WD systems cannot be detected by ground-based GW detectors such as the Advanced LIGO or
the future Einstein Telescope. However, they are appropriate targets for space-based interferometers like LISA,
TianQin (\citealt{Wang+etal+2019}) or Taiji (\citealt{Ruan+etal+2019}). Especially, it is found that LISA can 
essentially detect the GW emission from almost all SNe Ia within a distance of 20\,Mpc if they are triggered by 
the double-degenerate mechanism. As a result, LISA will be powerful tool to examine the trigger mechanism of SNe Ia. 

Many double WD binaries have been discovered in our Galaxy. Thus the merging of binary WDs will undoubtedly 
happen in our local Universe. Considering that the actual detecting distance of LISA  can be
significantly larger than 20\,Mpc, we believe that LISA will be able to detect plenty of merging WDs in the 
future. It is interesting to note that the luminosity distances of these chirping GW sources 
can be directly measured through GW observations themselves (\citealt{Schutz+1986,Messenger+Read+2012}). 
Additionally, the masses of each WD can also be measured. It is expected that LISA observations of future 
SNe Ia will not only help to examine the double-degenerate model of SNe Ia, but also provide valuable
information on the distances so that SNe Ia could act as more precise standard candles.

Traditionally, SNe Ia could only be studied based on multi-wavelength electro-magnetic 
observations (\citealt{Tutukov+Fedorova+2007}), assisted by some theoretical calculations  
on the chemical outcome during the bursting process (e.g. \citealt{Liu+etal+2018,Isern+Brav+2018}). 
As a result, it is hard to draw any firm conclusions on the progenitors. 
Gravitational wave can work as a completely new messenger for understanding the nature of SNe Ia.
LISA can hopefully help make firm constrains on this issue. It may lead the cosmology study 
into a new era (\citealt{Wang+etal+2003}). 
If the double-degenerate scenario was found correct, or the mechanism contributed at least a 
fraction of SNe Ia, then previous cosmology results based
on the single-degenerate scenario would be markedly modified.

\begin{acknowledgements}
  We acknowledge the anonymous referee for valuable suggestions that lead to an 
  overall improvement of this study.
  We thank Jinjun Geng for useful discussion and help.
  This work was supported by the National Natural Science Foundation of China (Grant No. 11873030),
  and by the Strategic Priority Research Program of the Chinese Academy of Sciences (``multi-waveband
  Gravitational Wave Universe'', Grant No. XDB23040000).
\end{acknowledgements}

\label{lastpage}

\end{document}